\documentclass[preprint2]{aastex}
\usepackage[dvips]{epsfig,color}

\newcommand{\eg}{{\it e.g.,}}
\newcommand{\ie}{{\it i.e.,}}
\newcommand{\etal}{{\it et al.}}
\newcommand{\cns}{CNS }
\newcommand{\ab}{AB }
\newcommand{\sersic}{S\'{e}rsic }

\shorttitle{Scalelengths in Halos}
\shortauthors{Barnes \etal}

\begin{document}

\title{Scalelengths in Dark Matter Halos}
\author{Eric I. Barnes, Liliya L. R. Williams}
\affil{Department of Astronomy, University of Minnesota, Minneapolis,
MN 55455}
\email{barnes@astro.umn.edu}
\email{llrw@astro.umn.edu}
\author{Arif Babul\altaffilmark{1}}
\affil{Department of Physics and Astronomy, University of Victoria,
BC, Canada}
\email{babul@uvic.ca}
\author{Julianne J. Dalcanton\altaffilmark{2}}
\affil{Astronomy Department, University of Washington, Box 351580,
Seattle, WA 98195}
\email{jd@astro.washington.edu}

\altaffiltext{1}{Leverhulme Visiting Professor,
Universities of Oxford and Durham}
\altaffiltext{2}{Alfred P. Sloan Foundation Fellow}

\begin{abstract}

We investigate a hypothesis regarding the origin of the scalelength in
halos formed in cosmological N-body simulations.  This hypothesis can
be viewed as an extension of an earlier idea put forth by Merritt and
Aguilar.  Our findings suggest that a phenomenon related to the radial
orbit instability is present in such halos and is responsible for
density profile shapes.  This instability sets a scalelength at which
the velocity dispersion distribution changes rapidly from isotropic to
radially anisotropic.  This scalelength is reflected in the density
distribution as the radius at which the density profile changes slope.
We have tested the idea that radially dependent velocity dispersion
anisotropy leads to a break in density profile shape by manipulating
the input of a semi-analytic model to imitate the velocity structure
imposed by the radial orbit instability.  Without such manipulation,
halos formed are approximated by single power-law density profiles and
isotropic velocity distributions.  Halos formed with altered inputs
display density distributions featuring scalelengths and anisotropy
profiles similar to those seen in cosmological N-body simulations.

\end{abstract}


\section{Introduction}

Some of the earliest studies to lay the foundation of galaxy formation
have relied on analytical methods to follow the collisionless
evolution of dark matter \citep[\eg][]{gg72,g75,g77} and lead to pure
power-law density distributions.  Several other studies have built on
this early analytical framework.  \citet{fg83} and \citet{b84} have
found similarity solutions for the spherical collapse of cosmological
perturbations that produce nearly pure power-law density profiles.
The halos in these studies are populated solely by radial orbits.
\citet{rg87} have introduced nonradial motion, but the resulting
density profiles are still well-approximated by a single power-law
over at least two orders of magnitude in radius, with noticeable
deviations only on small [$\log{(r/r_{\rm vir})} \lesssim -2$] and
large [$\log{(r/r_{\rm vir})} \gtrsim 0$] scales ($r_{\rm vir}$ is the
virial radius).

As computing power has increased, another approach has been adopted to
investigate cosmological structure formation, N-body simulations.
These studies do not suffer from the same limitations of adopted
approximations that are inherent in the analytically-based work,
however such simulations do rely on other approximations, like
gravitational softening.  Given appropriate initial conditions,
cosmological N-body simulations (CNS) can follow structure formation
through non-linear development and investigate the impact of
hierarchical merging \citep[\eg][]{dc91,nfw96,nfw97,m98,m99,b01,p03}.
These studies agree that the equilibrium dark matter halos that form
in \cns have nearly universal density profiles that are nothing like
single power-laws.  These halos generically have logarithmic slopes
$\gamma=-d\log{\rho}/d\log{r}$ that become larger with increasing
radius; $\gamma \approx 1$ at one-hundredth of a virial radius and
$\gamma \approx 3$ near the virial radius.  There remain, however,
disagreements regarding the exact values and behavior of $\gamma$,
especially at small radii.  

The fitting function that has been widely used to characterize the
density profiles of \cns halos, the NFW profile \citep{nfw96}, has an
explicit scalelength that divides the inner and outer behaviors.  The
NFW profile $\gamma$ asymptotes to 1 as $r\rightarrow 0$ and 3 as
$r\rightarrow \infty$.  Recently, \citet{n04} have discussed another
fitting function to describe the density profiles of higher resolution
\cns halos.  Unlike the NFW profile, $\gamma$ of the new function does
not approach asymptotic values at small and large radii and does not
have the same well-defined scalelength as the NFW profile.  Instead,
$\gamma$ of this new profile changes continuously with radius,
implying that there are no regions in which the density profile
behaves like a power-law.  At the same time, another recent study
\citep{d05} finds that the $r\rightarrow 0$ asymptotic power-law
behavior persists in their high resolution CNS.  Given this
disagreement, it is unclear whether or not \cns halos actually have
preferred scalelengths.   However, the NFW profile does provide a
reasonable description of density distributions for galaxy and cluster
scale \cns halos \citep{m05}.  For the sake of brevity, we will refer
to the radius at which $\gamma=2$ as the scalelength.

Despite these fairly minor disagreements regarding the behavior of
$\gamma$, \cns halos (as a class) appear qualitatively similar.  This
similarity does not extend to the analytically-based (AB) halos
described above.  Like previous studies that have sought to understand
the differences between \cns and \ab halos
\citep[\eg][]{afh98,l00,n01}, we are investigating how the input
physics differs between the methods and whether or not such physics
can explain the differences.  We find evidence that a phenomenon
related to the radial orbit instability (ROI) can account for the
differences as well as shed some light on the universality and
scalelengths of \cns halos.

A brief introduction to the ROI is given in \S\ref{roisec}.  We lay
out the hypothesis, drawing links between the ROI, \cns halos, and \ab
halos, in \S\ref{hyp}.  The demonstration of the hypothesis relies on
evolving dark matter halos semi-analytically and is based on the
method presented in \citet{rg87}.  Our modifications to this method
have been fully discussed in \citet{w04}.  Briefly, this method relies
on spherical symmetry and ignores complications due to merging.  This
simplicity provides us with a degree of control not present in CNS.
At the same time, inclusion of secondary perturbations allows us to
approximate the correct evolution of an isolated dark matter halo.
Details of the method and results of testing the hypothesis are
presented in \S\ref{test}.  We discuss the link between the ROI and
the shapes of \cns halo density profiles in \S\ref{link}.  The final
section summarizes and presents the conclusions drawn from this work.

\section{Overview of the Radial Orbit Instability}\label{roisec}

Early analytical and N-body studies have found that equilibrium
spherical systems composed of purely radial orbits are unstable to
forming bars or triaxial systems \citep[][\S5.2]{p81,fp84,bt87}.
Although there is no general theory explaining the ROI, \citet{pp87}
argue that if a spherical system develops a small $m=2$ distortion,
precession of radial orbits can draw them to reinforce the distortion,
thereby increasing the distortion and leading to bar formation.
Whatever the exact mechanism is, it is clear that nonradial forces
play a key role, as we will discuss in the next section.

\citet{ma85} have further investigated the ROI by creating equilibrium
systems with initial anisotropy ($\beta=1-\sigma_{\phi}^2/\sigma_r^2$)
profiles described by,
\begin{equation}\label{anisoeq}
\beta(r)=\frac{r^2}{r^2_a+r^2},
\end{equation}
where $r_a$ is referred to as the anisotropy radius \citep{m85}.
These profiles are completely isotropic ($\beta=0$) for
$r_a/r_{1/2}=\infty$, where $r_{1/2}$ is the initial half-mass radius.
For smaller $r_a/r_{1/2}$ values, the anisotropy increases from the
center outwards.  At the limit $r_a/r_{1/2}=0$, the entire model is
composed of radial orbits.  \citet{ma85} list several fundamental
results of the ROI.  First, an initially spherical system that
undergoes the ROI transforms into a nearly prolate bar with
long-to-short axis ratio $\approx 2-2.5$.  Second, for the anisotropy
profile of Equation~\ref{anisoeq} there appears to be a fairly
distinct border between systems that form a significant bar
($r_a/r_{1/2} \lesssim 0.2$) and those that remain relatively
spherical ($r_a/r_{1/2} \gtrsim 0.3$); however, other anisotropy
profiles produce different stability criteria.  Third, the global
anisotropy measure $2T_r/T_t$ (where $T_r$ and $T_t$ are the kinetic
energies associated with radial and tangential motions, respectively)
is not a universal arbiter of whether or not a system is unstable;
larger values of this ratio indicate instability, but there does not
appear to be a definitive demarcation value \citep{ps81,bgh86}.
The \citet{ma85} work also highlights that the stability of any
equilibrium system depends (at a minimum) on the anisotropy
distribution, with centrally isotropic systems being more stable than
those with radially anisotropic cores.  In brief, collisionless
systems will undergo the ROI if a sufficient fraction of orbits are
predominantly radial.

While we have so far discussed the ROI in equilibrium systems,
collisionless collapses also appear to relate to the ROI.  The link
with the ROI is that these cold collapses produce mostly radial
motion, a condition favorable to the onset of the ROI.  We compare the
results of noncosmological and cosmological collapses and discuss the
signature of the ROI in the next section.

\section{Collapses \& the Radial Orbit Instability}\label{hyp}

Studies of cold collisionless collapses in noncosmological settings
\citep{va82,m84,am90,tbva05} report that the end results of collapses
are good descriptions of observed surface brightness profiles of
elliptical galaxies.  For example, the collapses presented in
\citet{va82} result in mildly nonspherical systems with
two-dimensional projected density profiles that are well-approximated
by the \sersic function $\ln{(\Sigma/\Sigma_e)}=-b_n[(R/R_e)^{1/n}-1]$
\citep{s68}, with $n \approx 4$ like a de Vaucouleurs profile.  These
collapse products also display characteristic anisotropy profiles; the
inner regions are isotropic and transitions to radial anisotropy
occur at larger radii.  \citet{ma85} also point out that such
collapses are less centrally concentrated than purely radial
collapses.  We extend and focus these thoughts by specifically
investigating the link between density scalelengths and monotonic
variations in velocity anisotropy.

While some choices of initial conditions in the previously mentioned
studies have an eye towards describing cosmological situations, their
main goals relate to understanding elliptical galaxies.  Two studies
directly deal with the differences between cosmological and
noncosmological collapses.  \citet{k91} argues that the ROI does not
occur in cosmological settings because there is no correlation between
final shape and initial $2T/W$ value.  However, this study also
reports that the final states of collapses have density profiles that
are well-approximated by a Jaffe distribution \citep{j83}, which is
similar to a de Vaucouleurs (\sersic $n=4$) profile when projected to
two dimensions.  More importantly, it is stated that the equilibria
are supported by anisotropic velocity dispersions, like the results of
the noncosmological studies.  \citet{cm95} find that realistic
cosmological collapses do not succumb to the bar instability, but,
like \citet{k91}, the final density profiles are approximated by a
Jaffe distribution (again, similar to an $n=4$ \sersic profile).
Unfortunately, this study focuses solely on the final shapes of the
collapses and does not present any kinematic information to allow a
comparison of anisotropy profiles.  Beyond these two studies, it is
well-reported that halos formed in \cns tend to appear similar to the
products of noncosmological collapse simulations; \ie\ they appear
prolate spheroidal/triaxial \citep[\eg][]{cl96,swh04} and have
monotonically increasing anisotropy distributions \citep{cl96,h99}.
All of these points lead us to conclude that cosmological collapse
simulations result in systems that have the same qualitative density
and anisotropy profiles as those from noncosmological studies.

The final states of these collapses are similar, but have the systems
undergone the ROI?  From the literature, it appears that most
researchers consider bar formation to be the signpost of the
instability.  The \citet{k91} and \citet{cm95} studies suggest that
the ROI is absent in cosmological situations due to the lack of bars.
\citet{am90} refer to the merely slightly nonspherical results in
\citet{va82} as stable.

We propose that bar formation may not be the only sign of the ROI, but
possibly only the most flagrant.  The end results of collapses like
those in \citet{va82}, where no strong bar is formed, display a less
drastic result of the ROI.  The equilibrium states resulting from
collisionless collapses have similar anisotropy profiles; anisotropy
increases (becomes more radial) with radius.  This is in line with the
discussion in \citet{ma85} which regards systems with isotropic
central regions as more stable than those with radially anisotropic
central regions.  From this point of view, the ROI changes the mostly
radial initial velocity distribution into one that has an isotropic
core surrounded by a radial ``mantle''.  As we continue this
discussion, we will use the term ``mROI'' (mild aspect of the ROI) to
include the onset of the anisotropy profile and possible mild
triaxiality and to distinguish it from the usual (and more extreme)
bar-formation criterion.

\begin{figure}
\plotone{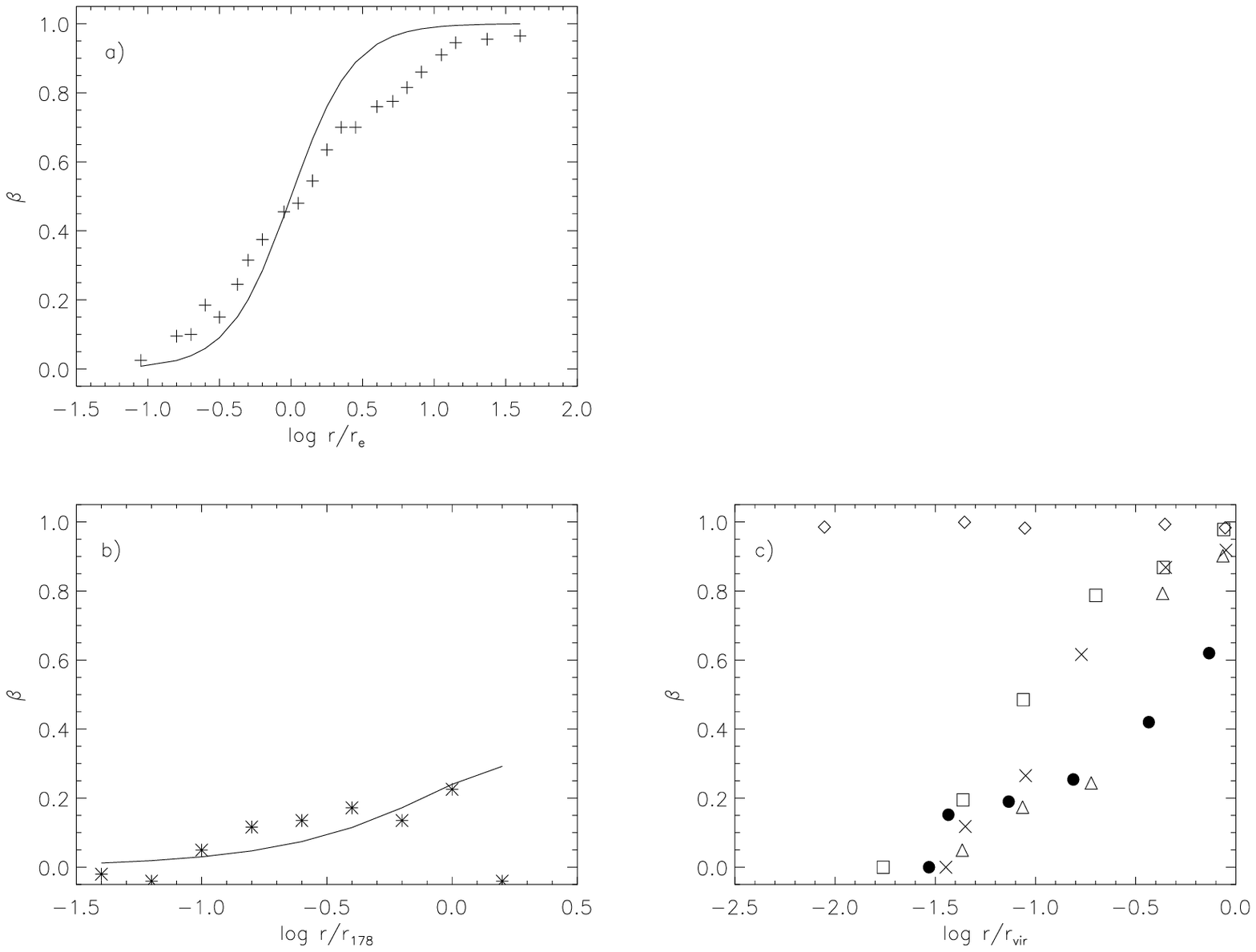}
\figcaption{Plots showing the behavior of the anisotropy parameter
$\beta$ for various simulations.  a) The plus symbols denote the
results of \citet{va82}.  The line is the initial anisotropy
distribution utilized in \citet{ma85} (see Equation~\ref{anisoeq}).
b) The asterisks mark the values resulting in the $n=-2$ simulation of
\citet{cl96}.  This line is the distribution suggested in \citet{c97}.
c) The following symbols represent the various models described in
\citet{h99}; diamonds -- Model I (purely radial), triangles -- Model
II, squares -- Model III, crosses -- Model IV, filled circles -- Model
V.
\label{bplot}}
\end{figure}

Figure~\ref{bplot} shows anisotropy profiles that can be constructed
from information in several previous studies.  Panel a) shows the
results of \citet{va82} (plus symbols) and the analytical expression
utilized by \citet{ma85}, Equation~\ref{anisoeq}.  In the figure, we
have set $r_a=r_e$, where $r_e$ is the effective radius of the
corresponding de Vaucouleurs profile.  Note that the analytical
expression is not a fit to the data points, it merely illustrates the
initial anisotropy profile utilized in the ROI study of \citet{ma85}.
Panel b) shows the results from the ``$n=-2$'' (here, $n$ refers to
the power spectrum $P(k)\propto k^n$, not the \sersic profile)
simulation of \citet{cl96} (asterisks) as well as the analytical
expression adopted by \citet{c97},
\begin{equation}
\beta=\beta_{\rm m}\frac{4r/r_{178}}{4+(r/r_{178})^2},
\end{equation}
where $r_{178}$ is the virial radius and we have chosen $\beta_{\rm
m}=0.3$.  This expression is intended to fit the data points from
equilibrium systems.  The last panel shows the results of the
simulations of \citet{h99}.  Admittedly, the amplitude of the
anisotropy in the \citet{cl96} result (panel b) is much smaller than
the other cases, but the monotonically increasing nature is still
present.  The mROI anisotropy profile is found generically in
collapses.

We find support for a link between the mROI and the presence of
scalelengths in \cns halos looking closely at the information from
\citet{h99}.  This work has investigated 5 models of isolated dark
matter halo formation.  Four of these have simple initial conditions
that were chosen to highlight the impact of differing amounts of
initial random motion.  The fifth represents a standard NFW type halo.
Their Model I is especially interesting since it has no initial
tangential motions and tangential forces are not allowed, essentially
forcing radial collapse.  The resulting halo is described by a
constant radial anisotropy ($\beta=1$, diamonds in
Figure~\ref{bplot}c) and a single power-law density profile $\rho
\propto r^{-2}$.  Once this restriction on tangential motion is lifted
(Models II-IV have increasing amounts of initial random motion and
allow nonradial forces), the resulting anisotropy profiles show
basically isotropic cores with $\beta$ increasing with radius
(Figure~\ref{bplot}c), evidence that the mROI is present in these
cases.  The corresponding density profiles resemble NFW profiles, \ie\
a scalelength is introduced.  While generic equilibria do not have a
unique relation between $\beta$ and $\gamma$, \citet{hm04} point out
that a wide variety of \cns halos appear to have a specific link
between the anisotropy distribution and $\gamma$.  In \S\ref{link}, we
further discuss this connection and the specific forms of the density
and anisotropy profiles.

These previous studies provide evidence of two things.  One, the mROI,
and its consequent anisotropy profile that has a distinct scalelength
dividing an isotropic core from a radial mantle, is ubiquitous in
collapses\footnote{We note that this evidence cannot be considered
conclusive as other studies have found that different mechanisms, most
notably density inhomogeneities, can also lead to similar outcomes
\citep[\eg][]{va82,l91}.}.  Two, this scalelength is linked to (if not
the same as) the scalelength of the resulting density distributions.
These points are the basis of the hypothesis that the mROI a) is the
key physics missing from previous \ab studies of dark matter halo
formation and b) leads to density profiles with scalelengths.  We will
now proceed to describe and discuss our testing of this hypothesis.

\section{Demonstrating the Link Between Anisotropy \& Density
Profiles}\label{test}

We are not interested in modeling the mROI.  What we want to do is
mimic the key result of the instability, the anisotropy profile, by
suitable alterations of inputs.  This approach will let us test the
root of the hypothesis, the impact of the anisotropy distribution on
the resulting density distribution.

To create the halos we will investigate, we follow the procedure
outlined in \citet{rg87} and \citet{w04}.  Initially thin
spherical shells expand with the Hubble flow until they reach their
turn-around radius, at which time they collapse onto the mass
interior.  At the moment a shell begins to collapse, the integrated
effects of secondary perturbations are introduced: a shell, which can
be thought of as a mass of particles, is given a velocity dispersion.
This can also be pictured as giving the shell a thickness, with apo-
and pericentric distances.  At the same time, the perturbation
velocity is oriented randomly, introducing angular momentum for that
shell.  The perturbation velocities calculated per \citet{rg87} are
actually RMS values.  Therefore, we follow \citet{w04} and randomly
choose the magnitude of the velocity from the Maxwell speed
distribution (truncated at 4 standard deviations) with the RMS value
from \citet{rg87}.  Since our method relies on choosing random
magnitudes and directions, for the rest of this paper we will average
over some number (usually 20) of halos and refer to the average
quantities as belonging to a single halo.

\begin{figure}
\plotone{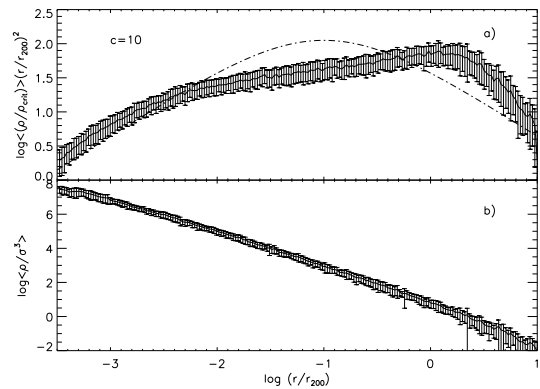}
\figcaption{Standard halo spatial and phase-space density proxy
distributions.  a) Multiplying the spatial density by $r^2$ enhances
the deviations from that power-law.  The dash-dotted line is an NFW
profile with a concentration of 10.  b) The phase-space density proxy
distribution.  The 1-$\sigma$ error bars are shown and the
normalization is arbitrary in both panels.
\label{stdden}}
\end{figure}

\begin{figure}
\plotone{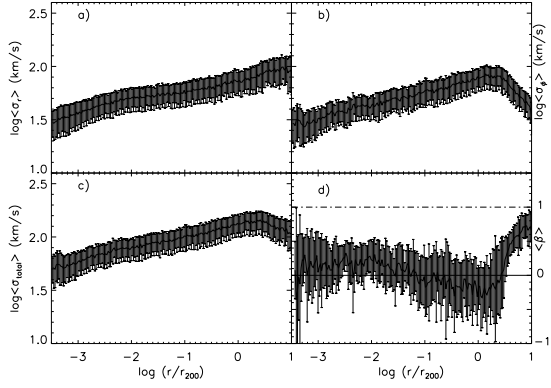}
\figcaption{Standard halo velocity distributions.  a)  The radial
velocity dispersion profile.  b)  The one-dimensional tangential
velocity dispersion profile.  c)  The total velocity dispersion
profile.  d) The anisotropy profile.  The dash-dotted line shows the
anisotropy of purely radial orbits.  The solid line is for isotropic
orbits.  The 1-$\sigma$ error bars are shown in all panels. 
\label{stdsig}}
\end{figure}

The benefit of the \citet{w04} formalism is that it allows us to
impose an arbitrary velocity dispersion profile and then study the
halos that result from the subsequent collapse.  However, before we
explore variations in the anisotropy profile, we first establish a
``standard'' halo to use as a benchmark.  By construction, it will be
similar to the halos that have been created in \citet{rg87} and
\citet{w04}.  The spatial density and phase-space density proxy
$\rho/\sigma^3$ profiles of the standard halo are shown in
Figure~\ref{stdden} (as usual, $r_{200}$ is the radius at which the
halo density is 200$\rho_{\rm crit}$).  The top panel highlights the
deviations from $\rho \propto r^{-2}$.  Over a substantial radial
range ($-2.5 \lesssim \log{(r/r_{200})}\lesssim 0.0$), the density is
well described by a single power-law $\rho \propto r^{-1.8}$.  This
agrees well with the halo described in \citet[][see their Figure
11]{rg87}.  We show the $\rho/\sigma^3$ distribution for the standard
halo in the bottom panel.  This distribution is nicely approximated by
a power-law with exponent $\approx -2$ over a wide radial range
$-2.5<\log{(r/r_{200})}<0$ \citep{a05}.  We also present the velocity
distributions in Figure~\ref{stdsig}.  The radial, one-dimensional
tangential, and total velocity dispersion profiles along with the
anisotropy distribution are shown.  Note that the velocities are
basically isotropic over the entire range (the upturn outside
$r_{200}$ is from infalling material).  Not surprisingly, given its
scale-free power-law properties, we find that the standard halo is a
poor match to those from cosmological N-body simulations, represented
in Figure~\ref{stdden} by a NFW profile of concentration 10
(dash-dotted line).

We now turn our attention to altering the parameters of the standard
halo by varying the velocities caused by secondary perturbations.  We
introduce a factor $\nu$, which is a function of initial comoving
radius $x$, to effect the desired changes.  Since we are interested in
mimicking the end result of the mROI, we introduce changes that
produce an anisotropy profile that is isotropic near the center and
becomes radially anisotropic with increasing radius.  

The prescription for doing this in \citet{w04} is a simple one; halve
the perturbation velocities in a specified comoving radial range
$(x_{\rm lo},x_{\rm hi})$ giving an inverted top-hat $\nu$
distribution,
\begin{equation}
\nu(x)=\left\{ \begin{array}
{r@{\quad : \quad}l}
1 & x<x_{\rm lo} \\ 0.5 & x_{\rm lo}<x<x_{\rm hi} \\ 1 & x>x_{\rm hi}.
\end{array} \right.
\end{equation}
In this prescription, the perturbation velocities are simply
multiplied by $\nu$.  The impact of this prescription is to decrease
the angular momentum imparted by perturbations in the specified radial
range.  This prescription also alters the total perturbation velocity,
thereby changing the energy imparted to the halo by the perturbations.
The end result is a halo that loses the power-law properties of the
standard halo and instead resembles an NFW profile \citep[see Figure 3
in][]{w04}.

\begin{figure}
\plotone{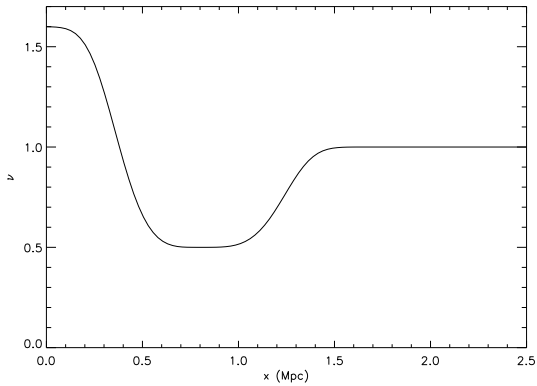}
\figcaption{The distribution of $\nu$ values versus initial comoving
radius in Mpc.  See \S\ref{test} for the utility of $\nu$.
\label{fdist}}
\end{figure}

A representative $\nu(x)$ distribution used in this study is shown in
Figure~\ref{fdist}.  The distance measure of the $x$-axis is the
initial comoving radius of a shell in Mpc.  The algebraic expression
is,  
\begin{equation}
\nu(x)=\nu^*(x)-[\nu^*(x)-d] e^{-(x-x_h)^n/2\omega^2},
\end{equation}
where $n$ and $\omega$ control the shape and width of the trough, $d$
is the depth of the trough, $x_h$ is the mid-point of the trough,
\begin{displaymath}
\nu^*(x)=0.5(\nu_0-\nu_{\infty})
(1-\tanh[2s(x-x_h)/(\nu_0-\nu_{\infty})])+\nu_{\infty},
\end{displaymath}
$2s/(\nu_0-\nu_{\infty})$ controls the slope of $\nu^*$ at $x_h$, and
$\nu_0$ and $\nu_{\infty}$ are the asymptotic values.  These
parameters give us considerable flexibility to explore the impact of
secondary perturbations on the final halos.  After some
experimentation, we have found that the interesting parameters to vary
are $\nu_0$, $\nu_{\infty}$, and $x_h$.  We fix the remaining
parameters; $n=4$, $\omega=0.4$, $d=0.5$, $s=20$.  In general,
$\nu_0>1$ and $\nu_{\infty}=1$.  If we simply multiply the
perturbation velocities by $\nu$, then near the center ($\nu>1$) we
introduce more angular momentum than in the standard halo.
Conversely, $\nu<1$ saps angular momentum relative to the standard
case.  Through various trials, this change from $\nu>1$ to $\nu<1$
appears essential to reproducing the mROI anisotropy profile, and we
point out that $\nu$ is simply a means of achieving this end.  We have
not determined the exact link between the behavior of $\nu$, which is
imposed only at a shell's initial turn-around point, and the physical
process of the mROI.  It is possible that the RMS perturbations
calculated according to \citet{rg87} are larger than those in \cns
because of the one-dimensional nature of this method.  If this is the
case, then $\nu<1$ is simply reducing the effects of exaggerated
secondary perturbations.  While further work may elucidate the
relationship between $\nu$ and the physics of the mROI, for the
purposes of this study, it suffices that this $\nu$ distribution gives
rise to an mROI-type anisotropy profile.

As it seems unlikely that the mROI changes the energies of particles,
we turn to a new prescription that leaves shell energies unchanged and
instead affects the orientations of the perturbation velocities.  The
prescription presented here (and shown in Figure~\ref{fdist}) uses the
following $\nu$ distribution parameters; $\nu_0=1.6$,
$\nu_{\infty}=1.0$, $x_h=0.8$ Mpc.  The magnitude of the perturbation
velocity remains unchanged from its value derived from the Maxwell
speed distribution, but $\nu>1$ increases the probability that the
velocity is oriented tangentially and vice versa.  We reiterate that
this prescription is intended solely to produce an mROI-type
anisotropy profile.  There is no direct link between $\nu$ and the
physics of the mROI, and so the correspondence between the final
profiles is bound to be inexact.

\begin{figure}
\plotone{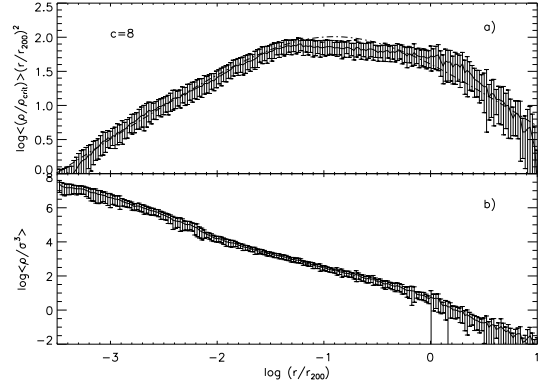}
\figcaption{The spatial (a) and $\rho/\sigma^3$ (b) density
distributions resulting from the prescription described in
\S\ref{test}.  The dash-dotted line in (a) is an NFW profile with a
concentration of 8.  The 1-$\sigma$ error bars are shown and the
normalization is arbitrary in both panels.
\label{try81den}}
\end{figure}

\begin{figure}
\plotone{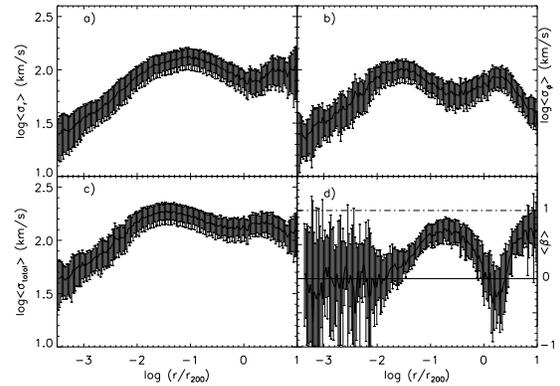}
\figcaption{The velocity distributions resulting from the prescription
described in \S\ref{test}.  The panels are the same as in
Figure~\ref{stdsig}.
\label{try81sig}}
\end{figure}

Figure~\ref{try81den} illustrates the resulting spatial density and
$\rho/\sigma^3$ profiles.  We can see that adopting this new
prescription has indeed provided a scalelength for the halo.  While we
have not set out to exactly recreate NFW halos, the similarity is
clear.  Some of the differences between the average and NFW profiles
is due to the fact that many individual halos are good approximations
to NFW profiles with different concentrations ($5\le c \le9$).
Averaging these together tends to smear the peaks into a broader
profile.  Looking at the distribution in panel (b), we see a profile
that is a decent approximation to a power-law over a couple orders of
magnitude.  It also appears to be a bit shallower (slope $\approx
-1.8$) than the standard halo's $\rho/\sigma^3$ profile and more in
line with the results of \citet{tn01} who find a slope of $-1.875$ for
three CNS halos.  Figure~\ref{try81sig} contains the same velocity
information as Figure~\ref{stdsig}.  The break in the power-law that
is present in the density has an analogous presence in the velocity
distributions.  Also, the anisotropy profile is familiar from the
preceding discussion of the mROI; \ie the velocities are isotropic
near the center with a fairly sharp transition to anisotropy.  This
result is in qualitative agreement with the discussion of collapses in
\citet{ma85}; the ROI reduces the central density relative to purely
radial collapses.

To this point, the discussion and figures presented in this section
are all based on a single set of parameters that determine the $\nu$
distribution.  We have done some parameter space exploration and have
found that the $\nu_0$ parameter can have an important impact on the
final profiles.  This is not very surprising, since it primarily
determines the impact of the isotropic core.  As $\nu_0$ is reduced
from 1.6 to 1, the inner $\gamma$ of the density profile changes from
$\approx 1$ to $\approx 1.3$.  Also, while we are not trying to
exactly reproduce NFW halos, we have noted that by eliminating the
introduction of random perturbation velocity magnitudes (\ie \ just
using the RMS values) we can make the agreement between our halos (not
shown) and NFW halos better.  We speculate that this occurs because it
provides ``colder'' initial conditions (in the sense discussed
regarding the ROI).  We also find that the anisotropy profile is
hardly affected if $\nu_{\infty}$ is varied between 1.0 and $d$ (the
effect of changing $\nu_{\infty}$ is to lower the $\nu$ value from 1.0
to $d=0.5$ over the range $1.5 \la x \le 2.5$, see
Figure~\ref{fdist}).

\begin{figure}
\plotone{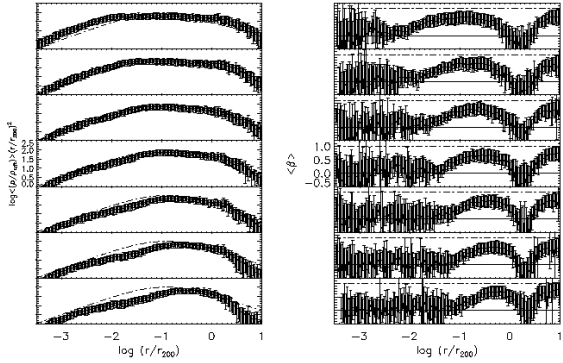}
\figcaption{Density (left column) and anisotropy (right column)
profiles for different values of $x_h$.  From top to bottom,
$x_h$=\{0.6, 0.7, 0.8, 0.9, 1.0, 1.1, 1.2\}.  The dash-dotted line in
the density profiles is an NFW profile with a concentration of 8.  The
ordinate values listed are the same for each panel in their respective
columns.  The solid and dash-dotted lines in the anisotropy profiles
mark isotropy ($\beta=0$) and total radial anisotropy ($\beta=1$),
respectively.  Each model is the average of 10 individual halos.
\label{suitec}}
\end{figure}

\begin{figure}
\plotone{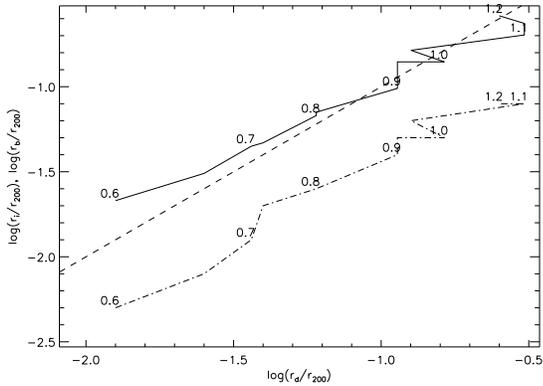}
\figcaption{The correlations between anisotropy radius $r_b$ (solid
line) and maximum isotropic radius $r_i$ (dash-dotted line) and the
density scalelength $r_d$.  The dashed line represents a one-to-one
correspondence and the numbers are the corresponding $x_h$ values.
\label{suitep}}
\end{figure}

A more thorough investigation has been made regarding the importance
of $x_h$, the mid-point of the $\nu$ trough.  This parameter directly
impacts the extent of the isotropic core present in the final
anisotropy profile.  In Figure~\ref{suitec}, we show density profiles
(left column) and the corresponding anisotropy profiles for 7 values
of $x_h$ surrounding the canonical value; from top to bottom
$x_h$=\{0.6, 0.7, 0.8, 0.9, 1.0, 1.1, 1.2\}.  We note that all the
density profiles have scalelengths, but the agreement with the NFW
form is within the error bars only for $x_h=\{0.7,0.8,0.9\}$.  More
importantly, these models highlight the correspondence between the
density scalelength $r_d$ and the anisotropy scalelength $r_b$,
defined here to be the radius at which $\beta=0.5$, as in
Equation~\ref{anisoeq}.  Figure~\ref{suitep} displays this correlation
(solid line) and also shows how $r_i$, the largest radius at which
$\beta=0$, changes with $r_d$ (dash-dotted line).  The density and
anisotropy scalelengths are roughly the same (within $\approx$ 20\%)
in these halos, further supporting the hypothesis that mROI induced
anisotropy profiles lead to scalelengths in density profiles.

\section{Relating the mROI and Shapes of N-body Halo Density
Profiles}\label{link}

In this section, we take up a discussion of the ``true'' shape of
N-body halos.  The preceding discussion has purposely ignored the
differences between the ``canonical'' NFW profile and other CNS halo
fitting functions \citep[\eg][]{n04}.  One must realize that our lack
of detailed understanding of the physics of halo formation has forced
us to approach halo structure from a very reactive posture; halos are
formed and then functions are chosen that resemble the resulting
density profiles.  Unfortunately, these functions are somewhat
arbitrary and do not, by themselves, lead to a deeper understanding of
halo dynamics.  The central idea of this study, that the mROI leads to
an anisotropy distribution that is linked to the density distribution,
can potentially provide a way around this arbitrariness.  A full
understanding of this physical mechanism can provide us with an
expected density profile for collapse equilibria.  In reality, the ROI
is not fully understood and so no {\it ab initio} prediction can be
made.  At present, we are left to continue the passive role of finding
the function(s) that best explain the simulation results and we now
discuss the density profile specifically.

The density profile that has long been found associated with both
noncosmological \citep[\eg][]{va82,l91} and cosmological
\citep[\eg][]{k91,cm95} N-body simulations is the de Vaucouleurs
(\sersic $n=4$) profile.  However, \sersic profiles are not very
similar to the NFW profile, the dominant density fitting function over
the past decade.  While NFW profiles do fit CNS halo density profiles
well, recent work by \citet{n04} suggests that the asymptotic behavior
of the NFW profile as $r \rightarrow 0$ and $r \rightarrow \infty$
does not fit density profiles as well as a function similar to the
\sersic form \citep[however, for a dissenting view see][]{d05}.
Additionally, \citet{dh01} and \citet{m05} have presented evidence
that the \sersic form is closer to the true shapes of halo density
profiles.  As the density profiles of the various ROI studies listed
in this paper also find \sersic (specifically de Vaucouleurs)
profiles, the N-body findings mesh nicely with the hypothesis that the
ROI is a key part of forming collapse equilibria.

Taking the link between the mROI and equilibrium density distribution
as established, we now speculate about the causal relationship between
them.  \citet{k04} point out that there are a wide variety of
equilibria available to halos with density profiles like those from
CNS.  This variety arises from different velocity distributions, but
we have shown that CNS halos display mROI-type anisotropy profiles
exclusively \citep[see also][]{hm04}.  We propose that this choice
between the equilibria is dynamically set by the mROI, which brings
about the anisotropy distribution that is associated with a \sersic
density profile.  As mentioned before, we cannot currently demonstrate
this association analytically, but \citet{tbva05}, following the work
of \citet{sb87} and \citet{mtj89}, discuss how the \sersic density
profile goes hand-in-hand with the isotropic core and radially
anisotropic mantle associated with the mROI.  They present a
distribution function meant to describe systems that have experienced
incomplete violent relaxation.  When this distribution function is
integrated over velocity space, the density profile is very similar to
the \sersic form.  Integrating this function over spatial coordinates,
one derives an anisotropy profile that has the characteristic mROI
shape.  In the context of this idea, the link between anisotropy and
density profiles is present regardless of initial conditions (\ie\
cosmological or noncosmological), but we speculate that such different
initial conditions most likely lead to differing \sersic $n$ values.
Returning to the ``true'' shape of N-body halos, we conclude that
S\'{e}rsic, rather than NFW, profiles are better descriptions of the
density profiles of CNS halos and that a fuller understanding of the
ROI will provide insight to the actual density profiles.

\section{Summary \& Conclusions}\label{sum}

There is an important difference between models of collisionless dark
matter halos created from analytically-based methods and those
resulting from cosmological N-body simulations.  Unless the input
physics \citep{afh98,l00,n01} or parameters are varied (as in
\S\ref{test}), the halos that result from analytical calculations tend
to be best described by a single power-law density profile and
approximately constant isotropic velocity dispersion anisotropy
profiles.  Cosmological N-body simulation halos tend to have density
distributions that steepen and anisotropy profiles that become more
radial with increasing radius.

We have explored the hypothesis, akin to the discussion in
\citet[][\S6]{ma85}, that the presence of scalelengths in cosmological
N-body simulations can be attributed to a mild aspect of the radial
orbit instability, which produces equilibrium halos with isotropic
cores surrounded by regions of radial anisotropy.  The hypothesis is
that it is this fundamental change in the character of the orbits
supporting the halo that leads to scalelengths.  Cosmological N-body
simulation halos incorporate the physics of the mild aspect of the
radial orbit instability and present such scalelengths.
Analytically-based halos do not have this instability ``built-in'' and
therefore lack the consequent scalelengths.

Utilizing an extension of the analytically-based method of
\citet{rg87}, we have investigated whether or not a preferred
scalelength can be introduced into a halo that would naturally have a
single power-law density distribution.  To mimic the impact of the
mild aspect of the radial orbit instability, we have artificially
altered the velocities introduced by secondary perturbations to be
more tangential near the center and more radial in the outer parts of
a halo.  This is an approximate way to include the effect of the
instability in a technique that does not explicitly include such
physics.  In our prescription, the magnitudes of the perturbation
velocities are left unchanged, only the directions are altered.  This
leaves the energy budget of the halo unchanged, which we argue is an
acceptable approximation to the instability.  The important result is
that the final halos have an anisotropy distribution reminiscent of
those from cosmological N-body simulation halos and a density profile
akin to an NFW profile.  An exploration of the parameter space of our
models shows that the presence of a scalelength is insensitive to the
relative sizes of the isotropic and radially anisotropic regions, and
the anisotropy radii of halos are approximately equal to their density
scalelengths.

If cosmological N-body simulations are environments in which the mild
aspect of the radial orbit instability plays a significant role (and
they appear to be so), two central features of such simulations have a
ready explanation.  One is the seeming universality of halo density
profiles which appears in both noncosmological and cosmological N-body
simulations as \sersic density profiles.  A physical mechanism, like
the mild aspect of the radial orbit instability, that generically acts
in collapses, even when mergers are absent, provides a simple
generator for such universality.  The outcome of the instability is a
variable anisotropy distribution with isotropic orbits in central
regions and radially anisotropic orbits in outer regions.  This
anisotropy distribution gives rise to density profiles with
scalelengths, the second defining feature of halos formed in
cosmological N-body simulations.

\acknowledgments
We thank an anonymous referee for a careful reading and insightful
comments that improved this paper.  This work has been supported by
NSF grant AST-0307604.  Research support for AB comes from the Natural
Sciences and Engineering Research Council (Canada) through the
Discovery grant program.   AB would also like to acknowledge support
from the Leverhulme Trust (UK) in the form of the Leverhulme Visiting
Professorship at the Universities of Oxford and Durham.  JJD was
partially supported through the Alfred P.\ Sloan Foundation.

\end{document}